\documentclass[twocolumn,times,trackchanges, dvipsnames]{aastex631}
\usepackage[utf8]{inputenc}
\usepackage{amsmath}

\begin{document}
\title{Rosette Nebula Outburst Gaia 24djk from the Young Stellar Object V557 Mon}

\author{Adolfo S. Carvalho}
\author{Lynne A. Hillenbrand}
\affiliation{Department of Astronomy; California Institute of Technology; Pasadena, CA 91125, USA}

\begin{abstract}
    A previously faint young stellar object (YSO), V557 Mon, rapidly brightened in late 2024 and is currently at least $\Delta G=3.3$ magnitudes brighter than its typical pre-outburst brightness. The ongoing outburst is identified in the Gaia Alerts system as Gaia24djk. We obtained a 1-2.5 $\mu$m spectrum of the object and find the spectrum is dominated by line emission and continuum excess consistent with rapid YSO accretion, similar to the star EX Lup during its outburst state. We speculate that the burst, which has not yet reached its peak brightness, may become an FU Ori outburst, which would be evidenced by the emission spectrum turning into an absorption spectrum.
\end{abstract}

\section{Introduction}
\label{sec:introduction}
The typical accretion rates of young stellar objects (YSOs) have, for decades, been measured to be much lower than is necessary for the objects to reach their main-sequence masses during their first few Myr. The solution to this timescale problem is that a large fraction of the mass is accreted during accretion outbursts of various types \citep{fischer_accretionVariabilityReview_2023ASPC}. The outbursts span a range of amplitudes and timescales, ranging from factors of a few to 1000s, and from days to decades, respectively. 

The most extreme accretion outbursts are those of FU Ori objects, in which the mass accretion rates increase by a factor of up to 10,000, and can stay that high for 100 years or more. The physical trigger(s) of FU Ori outbursts is(are) not yet known, though detailed studies of the time-evolution of the objects during and well after the peak of their initial outbursts, can help illuminate the physics of the trigger(s) \citep{Szabo_V1057cyg_2021ApJ, Carvalho_V960MonPhotometry_2023ApJ, Carvalho_HBC722_2024ApJ}. 

It has been challenging to catch new FU Ori objects during the early stage rising part of the outburst, as the disk is just beginning to overwhelm the magnetosphere. We present a near-infrared (NIR) spectrum of a high-amplitude outburst of the YSO candidate object V557 Mon \citep[designated Gaia24djk in the Gaia Alerts system\footnote{{\url{http://gsaweb.ast.cam.ac.uk/alerts/home}}},][]{GaiaMissionReference_2016A&A, Hodgkin_GaiaAlerts_2021A&A} and recommend further spectroscopic monitoring of the outburst.

\section{Data}
\begin{figure*}
    \centering
    \includegraphics[width=0.23\linewidth]{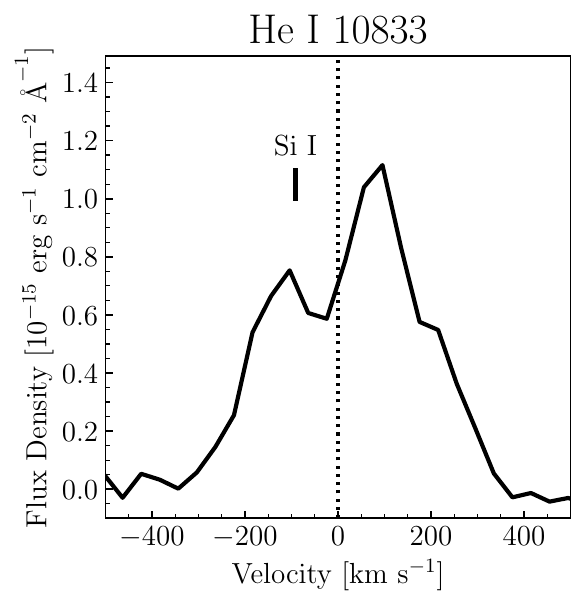}
    \includegraphics[width=0.236\linewidth]{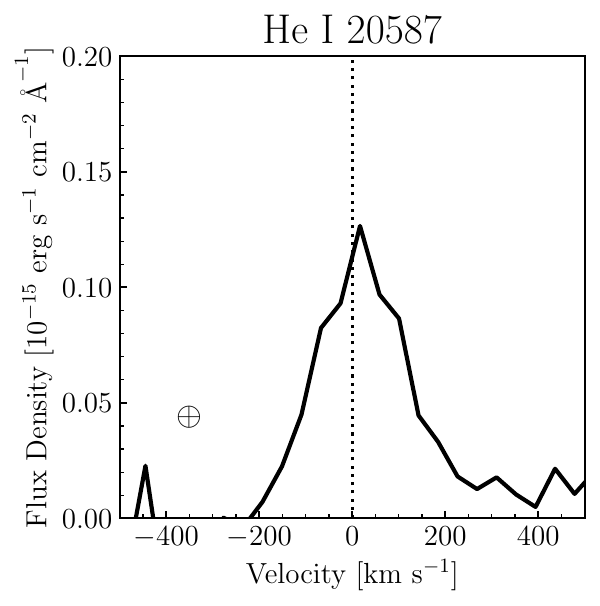}
    \includegraphics[width=0.23\linewidth]{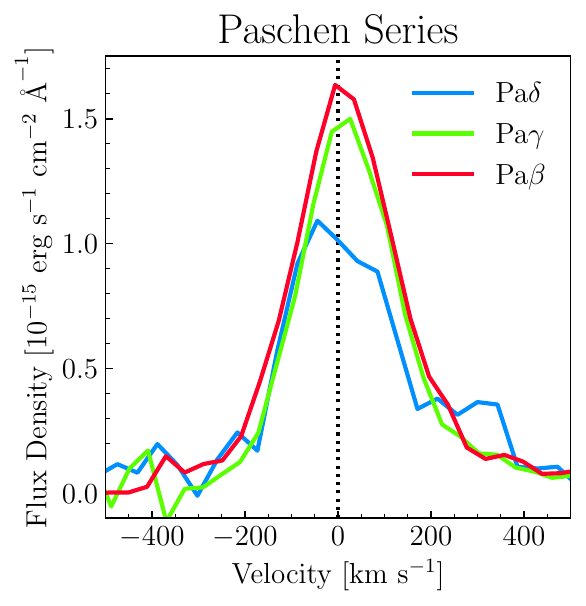}
    \includegraphics[width=0.23\linewidth]{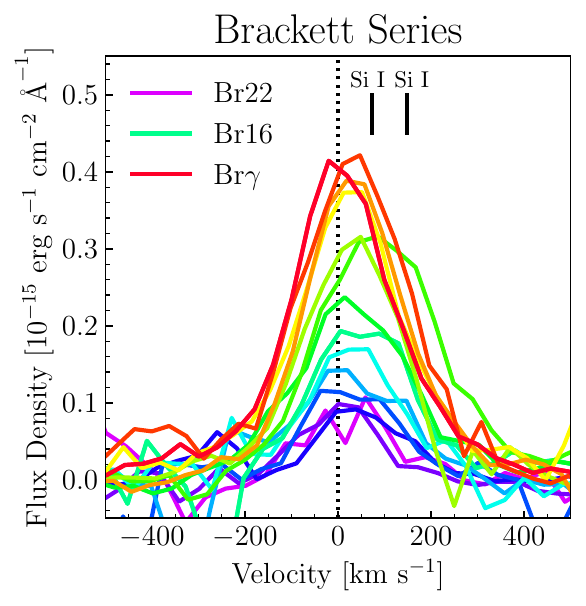}
    \includegraphics[width=\linewidth]{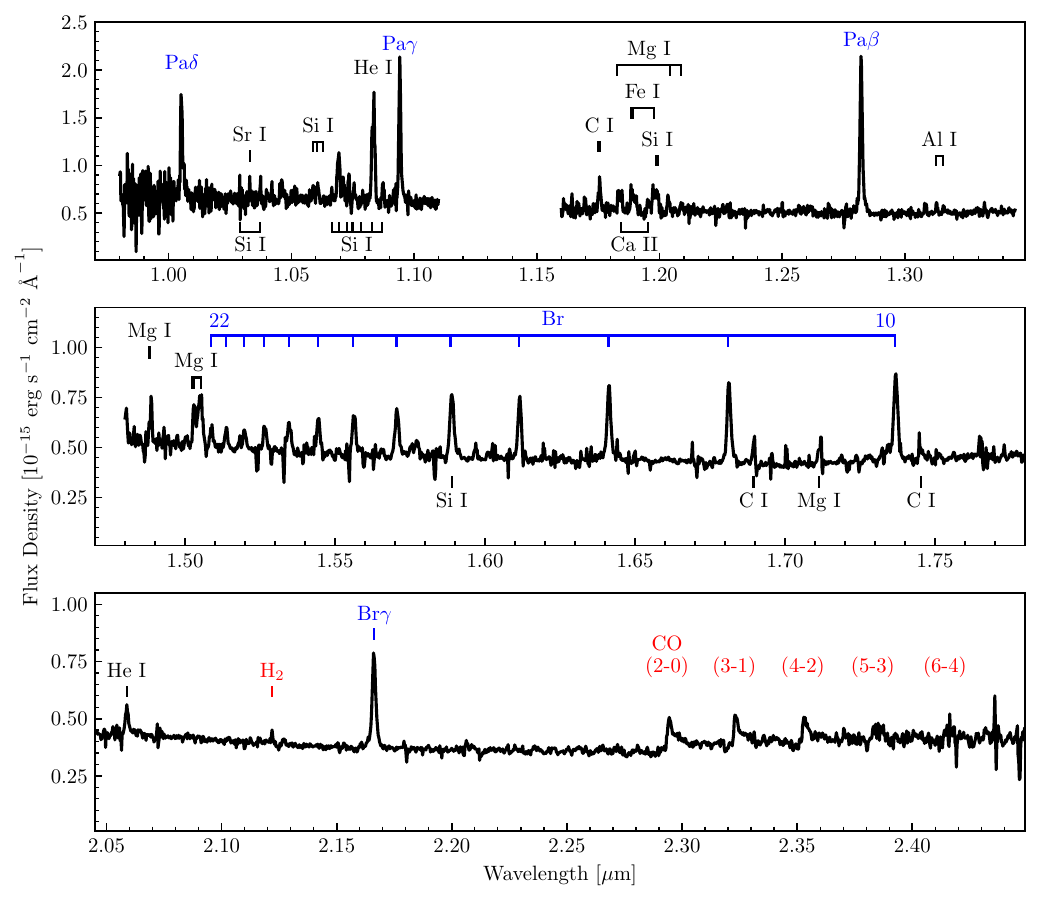}
    \caption{The NIR spectrum of Gaia24djk during the rapid rise phase of its outburst. \textbf{Upper Row:} Zoomed-in view of the continuum-subtracted line profiles of well-known features: \ion{He}{1} 10833 \AA, \ion{He}{1} 20587 \AA, the \ion{H}{1} Paschen series and the \ion{H}{1} Brackett series. The vertical ticks in the left and right panels mark locations of \ion{Si}{1} emission lines that contaminate the \ion{He}{1} and \ion{H}{1} profiles. The $\oplus$ marks an artifact due to heavy telluric absorption. \textbf{Lower Rows:} The full Palomar TripleSpec spectrum, showing several \ion{Si}{1}, \ion{Mg}{1}, and \ion{C}{1} lines in emission. The CO $\Delta \nu=2$ series starting at 2.29 $\mu$m can also be seen in emission, as well as the H$_2$ feature at 2.122 $\mu$m. }
    \label{fig:Spec}
\end{figure*}

We collected a 1.0-2.5 $\mu$m spectrum of the source using the TripleSpec instrument on the Palomar Observatory 200 in. Hale Telescope \citep{Wilson_TSpecPaper_2004SPIE}. The spectrum was taken on 16 January 2025 in four 300s exposures using an ABBA nod pattern, for a total exposure time of 1200s. The spectra were extracted, coadded, and telluric-corrected using $\mathtt{spextool}$ \citep{Cushing_spextool_2004PASP}. The final peak signal-to-noise ratio of the spectrum is 10/20/40/80 in Y/J/H/K bands. The spectrum is shown in Figure \ref{fig:Spec}.

\section{The YSO V557 Mon}
The progenitor of this outburst is identified as V557 Mon, a probable member of young cluster NGC 2244 in the Rosette Nebula \citep{Muzic_Rosette_NGC2244_2022A&A}. NGC 2244 has an estimated age of 2 Myr and is at a distance of 1440 $\pm$ 32 pc \citep{Muzic_Rosette_NGC2244_2022A&A}. 

The pre-outburst SED is consistent with a star/disk system having a photospheric temperature 3500-5500 K extincted by an $A_V$ of 1.5-3.5 mags. The SED also has both blue/UV and $K$ band excesses as is typical for Class I/II YSOs. The WISE \citep{Cutri_WISE_Reference_2012yCat} $W_1 - W_2 = 0.83$ and $W_2-W_3 = 3.23$ colors of V557 Mon pre-outburst also indicate a Class I/II classification, according to the WISE color-based YSO classification criteria adopted by \citet{Koenig_WISE_YSOClass_2014ApJ}. 

Prior to this outburst, there was flare-like burst of $\Delta r = 1.3$ mag detected in ZTF in February 2021. The burst was short-lived, lasting only 17 days. The object was observed by WISE at this time and the NEOWISE \citep{Mainzer_neowise_2011ApJ} time series photometry also shows the burst with a nearly colorless $\Delta W1 = 0.65$. 

\section{2024 outburst of V557 Mon}
The recent optical outburst began around 15 August 2024 and by 18 November 2024 the object had brightened by 3.3 magnitudes in the Gaia $G$ band, corresponding to a rise rate of 0.035 mag/day or 1.1 mag/month. The brightening is also seen in Zwicky Transient Facility \citep[ZTF,][]{Bellm_ZTFReference_2019PASP} photometry of the system, going from $g = 20$ ($r = 19.3$) to $g = 17.7$ ($r = 16.7$) by 14 November 2024. From our spectrum we see that the object is currently at least $\Delta J = 1.5$ magnitudes brighter than its pre-outburst state. 

The $g-r$ color increased from 0.7 to 1.0 during the rise. If the brightening were due to a change in the line-of-sight extinction to the system, it should have become bluer as it brightened. The relative lack of color, or slight reddening, might be expected if the disk is moving inward and contributing more to redder bands as it heats up, as would be the case for a developing FU Ori outburst. 

\section{outburst spectrum}

The spectrum of the outburst source Gaia24djk is similar to that of an EX Lup-type outburst \citep[e.g.,][]{Aspin_V2492Cyg_2011AJ, Kuhn_V1741Sgr_2024MNRAS}, dominated by atomic emission from \ion{H}{1}, \ion{He}{1}, \ion{Si}{1}, \ion{Mg}{1}, \ion{Fe}{1}, and \ion{C}{1}, and molecular emission from CO and H$_2$. The spectrum is extremely flat, implying the NIR continuum includes significant contribution from the viscously-heated accretion disk. 

The emission spectrum of the object is exquisite, with the entire Brackett $H$ band sequence from Br10 to Br22 clearly visible and the hot \ion{He}{1} 1.0833 and 2.058 $\mu$m lines both also readily apparent. It is uncommon (if not unprecedented) to see so many of the Brackett series in low mass YSOs. Furthermore, the 20-21 eV upper state excitation energy of the \ion{He}{1} features requires a relatively high temperature for the emission source \citep{NIST_ASD}. 

It is possible, therefore, that the current brightening of V557 Mon is the onset of a more extreme accretion event than a typical EX Lup-type burst. In future work, we will present a study of the object's evolution in the early stages of the burst and estimates of its physical properties. 

Photometric and spectroscopic monitoring in this early phase of the outburst will be critical to better understanding objects like this, whether the burst is of EX Lup or FU Ori type. 

\section{Acknowledgements}
\software{Astropy \citep{astropy_2013, astropy_2018, astropy_2022}, NumPy \citep{harris2020array}}

\bibliography{references}{}
\bibliographystyle{aasjournal}



\end{document}